\documentstyle[12pt,epsf]{article}
\newcommand{\noi}{\noindent}
\newcommand{\eq}{\begin{equation}}
\newcommand{\en}{\end{equation}}
\newcommand{\eqa}{\begin{eqnarray}}
\newcommand{\ena}{\end{eqnarray}}

\newcommand{\psb}{\overline{\psi}}
\newcommand{\psbps}{\overline{\psi} \psi}
\newcommand{\psbgps}{\overline{\psi} \gamma_5 \psi}

\newcommand{\vx}{{\vec x}}
\newcommand{\vy}{{\vec y}}
\newcommand{\eff}{e \! f \! f}
\newcommand{\mpr}{{\tilde m}}
\newcommand{\hg}{{\hat \Gamma}}

\newcommand{\aleq}{\mbox{}_{\textstyle \sim}^{\textstyle < }}
\newcommand{\ageq}{\mbox{}_{\textstyle \sim}^{\textstyle > }}
\newcommand{\ra}{\rightarrow}

\newcommand{\be}{\begin{equation}}
\newcommand{\ee}{\end{equation}}
\newcommand{\bea}{\begin{eqnarray}}
\newcommand{\eea}{\end{eqnarray}}

\newcommand{\np}{n^{\prime}}

\newcommand{\spr}{s^{\prime}}

\begin{document}
\hbox{}
\noindent  May 1996 \hfill DESY 95-141, HU Berlin--IEP--95/5
\begin{center}
\vspace*{1.5cm}

\renewcommand{\thefootnote}{\fnsymbol{footnote}}
\setcounter{footnote}{1}

{\LARGE On the chiral limit in lattice gauge theories 
with Wilson fermions}
\footnote{Work supported by the Deutsche
Forschungsgemeinschaft under research grant Mu 932/1-3 } \\

\vspace*{0.5cm}
{\large
A.~Hoferichter, 
V.K.~Mitrjushkin 
\footnote{Permanent address:
Joint Institute for Nuclear Research, Dubna, Russia}
and
M.~M\"uller-Preussker 
}\\
\vspace*{0.2cm}
{\normalsize
{\em Institut f\"{u}r Physik, Humboldt-Universit\"{a}t,
10099 Berlin, Germany}\\
}    

\vspace{1cm}
{\bf Abstract}
\end{center}
The chiral limit $~\kappa \simeq \kappa_c(\beta)~$ in lattice gauge
theories with Wilson fermions and problems related to near--to--zero
('exceptional') eigenvalues of the fermionic matrix are studied.  For
this purpose we employ compact lattice QED in the confinement phase.
A new estimator $~\mpr_{\pi}~$ for the calculation of the pseudoscalar
mass $~m_{\pi}~$ is proposed which does not suffer from 'divergent'
contributions at $\kappa \simeq \kappa_c(\beta)$.  We conclude that the
main contribution to the pion mass comes from larger modes, and
'exceptional' eigenvalues play {\it no} physical role.   The
behaviour of the subtracted chiral condensate $~\langle \psb \psi
\rangle_{subt}~$ near $~\kappa_c(\beta)~$ is determined.  We observe a
comparatively large value of $~\langle \psb \psi \rangle_{subt} \cdot
Z_P^{-1}~$, which could be interpreted as a possible effect of the
quenched approximation.

\section{Introduction}

As it is well known, chiral symmetry is broken explicitly in lattice
gauge theories with Wilson fermions as in QCD and QED \cite{wil}.
Presumably, it can be restored by fine--tuning the parameters in the
continuum limit.  If so, one can approach the continuum limit and 
chiral symmetry restoration along a 'critical' line $~\kappa_{c}(\beta)$.  
It is another question, whether on this line the chiral symmetry becomes
explicitly realized or spontaneously broken. 
In the continuum limit the lowest--lying state in the spectrum of the
Wilson fermionic matrix $~{\cal M}(\kappa ;U)~$ should have eigenvalue zero for
$~\kappa \ra \kappa_c(\beta )~$.  

For {\it nonzero} lattice spacing $~a \neq 0~$ the chiral symmetry
cannot be restored exactly. At the same time in the confinement phase
 the pion mass $~m_{\pi}~$
tends to zero in the limit $~\kappa \ra \kappa_c(\beta )~$ (the so--called
partial symmetry restoration).  The mechanism of this partial symmetry
restoration  is still not well determined.  
For example, one can not exclude that it is connected with the transition to
a parity violating phase where 
$~\langle \psb \gamma_5 \psi \rangle \neq 0$ \cite{aoki}.
If this transition is of second order then there will be a massless
pion at $~\kappa = \kappa_c(\beta)$.

Computations near the chiral transition line $~\kappa_c(\beta )~$
are notoriously difficult.  The first problem is that the matrix
inversion becomes very slow when $~\kappa~$ approaches 
$~\kappa_c(\beta)~$ and most of the known inversion methods fail 
very close to $~\kappa_c(\beta)$. Nevertheless, the conjugate gradient
method \cite{cg1,cg2} appears to be reliable even in the 'critical' region.
Another problem which arises in simulations of
lattice gauge theories with Wilson 
fermions is that near the chiral transition
very small ('exceptional')
eigenvalues $~\{\lambda_i \sim 0 \}~$ of the 
fermionic matrix $~{\cal M}(\kappa ; U)~$ appear which make it
practically impossible to approach $~\kappa_c(\beta)~$ without an 
enormous increase of statistics. 
The usual way to bypass this problem is to 
carry out calculations at $~\kappa$--values 
sufficiently below $~\kappa_c(\beta)~$ and then
to extrapolate the observables, e.g., hadron masses, 
to the 'critical' value $~\kappa_c(\beta)$. 

Apart from practical considerations the zero mode problem is connected
with the question of the mechanism of the chiral transition in lattice
theories with Wilson fermions. As it was already mentioned, zero
eigenvalues are expected to exist in the chiral limit in the continuum.
Within this context the question has to be answered, whether the
near--to--zero eigenvalues of $~{\cal M}(\kappa;U)~$ observed on the
lattice in the 'critical' region in finite volumes have physical
relevance or rather arise as an effect of the lattice discretization.

Configurations with extremely small eigenvalues $~\{\lambda_i \}~$
were first discovered in \cite{eig1,eig2}, and were called 'exceptional'
configurations.  In fact, very close to the transition 
point $~\kappa \simeq \kappa_c(\beta)~$ such configurations 
appear to be very 'normal'
(see, e.g. \cite{bkr,blls,qed1,qed2}).
In practice the appearance of these configurations can be used as an
indicator for approaching $~\kappa_c(\beta)$.

It is important to point out that the appearance of near--to--zero
eigenvalues $~\{\lambda_i \}~$ 
at $~\kappa \sim \kappa_c(\beta)~$ is {\it not}
a disease of the quenched approximation.
The fermionic determinant decreases the spread of small eigenvalues
but does not eliminate them totally \cite{eig2,bkr,qed1,qed2}.

It is the aim of this work to study the properties of a lattice 
gauge theory with Wilson fermions very close to the chiral
transition line $~\kappa_c(\beta )$.
In the confinement phase we study the behaviour of the pseudoscalar
pion mass $~m_{\pi}~$ and the subtracted chiral condensate
$~\langle \psb \psi \rangle_{subt}~$, which for Wilson fermions
is an order parameter of chiral symmetry breaking
in the continuum limit, near $~\kappa_c(\beta)$.
We employ compact lattice QED in the quenched approximation.
Compact lattice QED with Wilson fermions in the confinement 
phase possesses similar features near the chiral transition
$~\kappa \sim \kappa_c(\beta )~$ as nonabelian theories, and thus
can serve as a comparatively simple model of QCD in this study.
The cost of numerical calculations is much less  than for QCD.
We expect that our analysis is applicable to QCD, as well.

\section{Action and observables}

The standard Wilson lattice action
$ S_{WA}(U, {\bar \psi}, \psi)$ for $4d$ compact $~U(1)~$ gauge 
theory (QED) is

\eq
 S_{WA} = S_{G}(U) + S_{F}(U, {\bar \psi}, \psi) .
                                              \label{wa}
\en

\noi In eq.(\ref{wa}) $~S_{G}(U)~$ is the plaquette
(Wilson) action for the pure gauge  $~U(1)~$ theory

\eq
 S_{G}(U) =
\beta \cdot \sum_{P}
        \,  \bigl( 1 - \cos \theta_{P} \bigr) ~,
                                              \label{wag}
\en

\noi where $~\beta = 1/g^{2}_{bare}~$, and
$~U_{x \mu} = \exp (i \theta_{x \mu} ),
\quad \theta_{x \mu} \in (-\pi, \pi] ~$
are the field variables defined on the links $l = (x,\mu)~$.
Plaquette angles $~\theta_{P}  \equiv \theta_{x;\, \mu \nu}~$
are given by
$~\theta_{x;\, \mu \nu} =
  \theta_{x;\, \mu} + \theta_{x + \hat{\mu};\, \nu}
- \theta_{x + \hat{\nu};\, \mu} - \theta_{x;\, \nu} ~$.

The fermionic part of the action
$S_{F}(U, {\bar \psi}, \psi)~$ is

\eqa
S_{F}
& = & \sum_{x,y} \sum_{s,\spr=1}^{4}
{\bar \psi}_{x}^{s} {\cal M}^{s \spr}_{xy} \psi_{y}^{\spr}
\equiv  {\bar \psi} {\cal M} \psi~,
\nonumber \\ \nonumber \\
{\cal M} & \equiv & \hat{1} - \kappa \cdot Q(U) ,
\nonumber \\ \nonumber \\
Q^{s \spr}_{xy}
& = & \sum_{\mu}
\Bigl[ \delta_{y, x+\hat{\mu}} \cdot ( {\hat 1} - \gamma_{\mu})_{s \spr}
\cdot U_{x \mu} +
\delta_{y, x-\hat{\mu}} \cdot ( {\hat 1} + \gamma_{\mu})_{s \spr} \cdot
U_{x-\hat{\mu}, \mu}^{\dagger} \Bigr]~,
                                              \label{waf}
\ena

\noi where $~{\cal M}~$ is Wilson's fermionic matrix,
and $\kappa$ is the hopping parameter.

It is known, that at least in the strong coupling region for the
standard Wilson action
(in the confinement phase) the ordinary mass term and the Wilson mass term
cancel within the pseudoscalar mass $~m_{\pi}~$ at some
$~\kappa = \kappa_c(\beta)~$,
so that quadratic terms in the effective potential vanish for
the pseudoscalar field , and $\kappa_c \sim 0.25$ at 
$\beta=0$ \cite{kawa,kasm}.

In the weak coupling range  perturbative calculations
indicate that the mass of the fermion becomes equal to zero along the
line $\kappa_{c}(\beta )$ (for the free field theory $\kappa_{c}=0.125$)
\cite{kawa}.

We calculated the following fermionic observables

\eqa
\langle {\bar \psi} \psi \bigr \rangle
& = & \frac{1}{4V} \cdot
\langle \, \mbox{Tr} \Bigl({\cal M}^{-1} \Bigr) \rangle_{G}~;
\qquad
\langle {\bar \psi} \gamma_{5}  \psi \bigr \rangle
= \frac{1}{4V} \cdot
\langle \, \mbox{Tr} \Bigl( \gamma_{5} {\cal M}^{-1}\Bigr) \rangle_{G}~;
\nonumber \\
\nonumber \\
\langle \Pi \rangle
& = & \frac{1}{4V} \cdot \langle \mbox{Tr} \Bigl( {\cal M}^{-1}
\gamma_{5} {\cal M}^{-1} \gamma_{5} \Bigr) \rangle_{G}~,
                                      \label{pionnm}
\ena

\noi where $~\langle ~~ \rangle_{G}~$ stands for averaging over gauge
field configurations, and $~V=N_{\tau} \cdot N_s^3~$ is the number of sites.
Pseudoscalar zero--momentum correlators $~\Gamma (\tau )~$ are 
defined as follows

\eqa
\Gamma (\tau )
&=& - \frac{1}{N_s^6} \cdot \sum_{\vec{x},\vec{y}}~
\langle {\overline \psi} \gamma_5 \psi (\tau ,\vx ) \cdot
{\overline \psi} \gamma_5 \psi (0, \vy ) \rangle 
\nonumber \\
 \nonumber \\
& \equiv & \frac{1}{N_s^6} \cdot \sum_{\vec{x},\vec{y}}~
\langle \Biggl\{ \mbox{Sp}
\Bigl( {\cal M}^{-1}_{x y} \gamma_5 {\cal M}^{-1}_{y x} \gamma_5 \Bigr)
- \mbox{Sp} \Bigl( {\cal M}^{-1}_{x x} \gamma_5 \Bigr) \cdot
  \mbox{Sp} \Bigl( {\cal M}^{-1}_{y y} \gamma_5 \Bigr) \Biggr\} 
\rangle_{G}
\ena

\noi where ~~Sp~~ means the trace with respect to the Dirac indices.
These correlators  as well as the pion norm  $~\Pi~$ appear to be very
sensitive observables in the 'critical' region.
This can be understood by considering the spectral representation
of the fermionic order parameters.
Let $~f_{n} \equiv f_{n}(s,x)~$ be the eigenvectors of
$~{\cal M}~$ with eigenvalues $~\lambda_{n}~$,
and $~g_{n} \equiv g_{n}(s,x)~$ be the eigenvectors of
$~\gamma_{5} {\cal M}~$ with eigenvalues $~\mu_{n}~$ :

\eq
{\cal M} f_{n}
= \lambda_{n} \cdot f_{n}~,
\quad \quad
\gamma_{5} {\cal M} g_{n}
= \mu_{n} \cdot g_{n}~.
\en

\noi Then one can easily obtain a spectral representation
of the fermionic order parameters~:

\eqa
\langle {\bar \psi} \psi \bigr \rangle
& = & \frac{1}{4V} \Bigl \langle~\sum_{n} \frac{1}{ \lambda_{n}}~\Bigr
\rangle_G~,
\qquad
\langle {\bar \psi} \gamma_{5}  \psi \bigr \rangle
= \frac{1}{4V} \Bigl \langle~ \sum_{n} \frac{1}{ \mu_{n}}~\Bigr \rangle_G~,
\nonumber \\ \nonumber \\
\langle \Pi \bigr \rangle
& = & \frac{1}{4V} \Bigl \langle~\sum_{n} \frac{1}{\mu_{n}^2}~\Bigr 
\rangle_G~.
                                 \label{spectral}
\ena

\noi Evidently, an eigenstate of $~{\cal M}~$ with eigenvalue zero is
also an eigenstate of $~\gamma_{5} {\cal M}~$. So, the presence of
configurations which belong to zero eigenvalues of $~{\cal M}~$
also gives rise to poles in $~\Pi~$.
 For correlators one obtains

\eq
\sum_{\vec{x}} \mbox{Sp}
\Bigl( {\cal M}^{-1}_{x 0} \gamma_5 {\cal M}^{-1}_{0 x} \gamma_5 \Bigr)
|_{x_4=\tau }
= \sum_{n \np} \frac{1}{\mu_{n}}
\cdot b_{n \np}(\tau ) \cdot \frac{1}{\mu_{\np}} \cdot b_{\np n}(0) ~,
\en

\noi where

\eq
b_{n \np}(\tau ) \equiv
\sum_{\vx ,s} g^{\ast}_{n}(s,\vx ,\tau ) \cdot g_{\np}(s,\vx ,\tau )~.
\en

For further discussion on properties of the fermionic matrix
see, e.g., \cite{iiy,uka} and references therein.

\section{The pion mass in the chiral limit}
\subsection{Pseudoscalar correlators and the standard estimator 
$~m_{\pi}~$ near $~\kappa_c(\beta)~$ \label{sect_pm}}

Transfer matrix arguments suggest the following form of
the pseudoscalar correlator $~\Gamma(\tau)~$ (at least, for
$~\tau > 0~$) \cite{lues}

\eq
\Gamma(\tau) = A_{\pi} \cdot \Bigl[ e^{-m_{\pi} \cdot \tau}
+ e^{-m_{\pi} \cdot (N_{\tau} - \tau)} \Bigr] +
~~ \mbox{higher--energy states}~.
                      \label{transf}
\en

\noi The standard  choice of the estimator for the effective mass of the
pseudoscalar particle $~m^{\eff}_{\pi}(\tau ) \equiv m_{\pi}(\tau )~$ is

\eqa
\frac{\cosh m_{\pi}(\tau ) \left( \frac{N_{\tau}}{2} - \tau -1 \right) }
{\cosh m_{\pi}(\tau ) \left( \frac{N_{\tau}}{2} - \tau  \right) }
&=& \frac{\hg (\tau +1)}{\hg (\tau )}
                   \label{mass_stand}
\\
\nonumber \\
&=& e^{- m_{\pi}(\tau )}~
\quad \mbox{if} \quad m_{\pi}N_{\tau} \gg 1~,
\nonumber
\ena

\noindent where $\hg (\tau )$ is  the estimator of 
the pseudoscalar zero--momentum correlator

\eq
\hg (\tau ) = \frac{1}{n} \cdot \sum_{i=1}^{n} \Gamma_i (\tau ) ~,
                 \label{corr1}
\en

\noi and $~n~$ is the number of measurements.
The $~\tau$--dependence of the effective mass stems from the contribution
of higher--energy eigenstates of the transfer matrix.
With increasing $~\tau~$ the contribution of these higher--energy
states is expected to be suppressed, and the resulting plateau 
in the $~\tau$--dependence of the effective mass gives the
true mass $~m_{\pi}~$.

However this approach fails when one comes close to 
the chiral limit, i.e.,
when $~\kappa \rightarrow \kappa_c(\beta)~$. 
The well--known problem in both QED and QCD with
the calculation of $~m_{\pi}~$ 
(and other fermionic observables) is connected with extremely small
eigenvalues of the fermionic matrix $~{\cal M}~$ (and correspondingly
$~\gamma_5 {\cal M}$) which appear on a finite lattice at some 
$~\kappa^{\prime}(\beta ;N_s;N_{\tau}) \aleq \kappa_c (\beta)~$.

%
%
\begin{figure}[htb]
\begin{center}
\vskip -2.5truecm
\leavevmode
\hbox{
\epsfysize=570pt\epsfbox{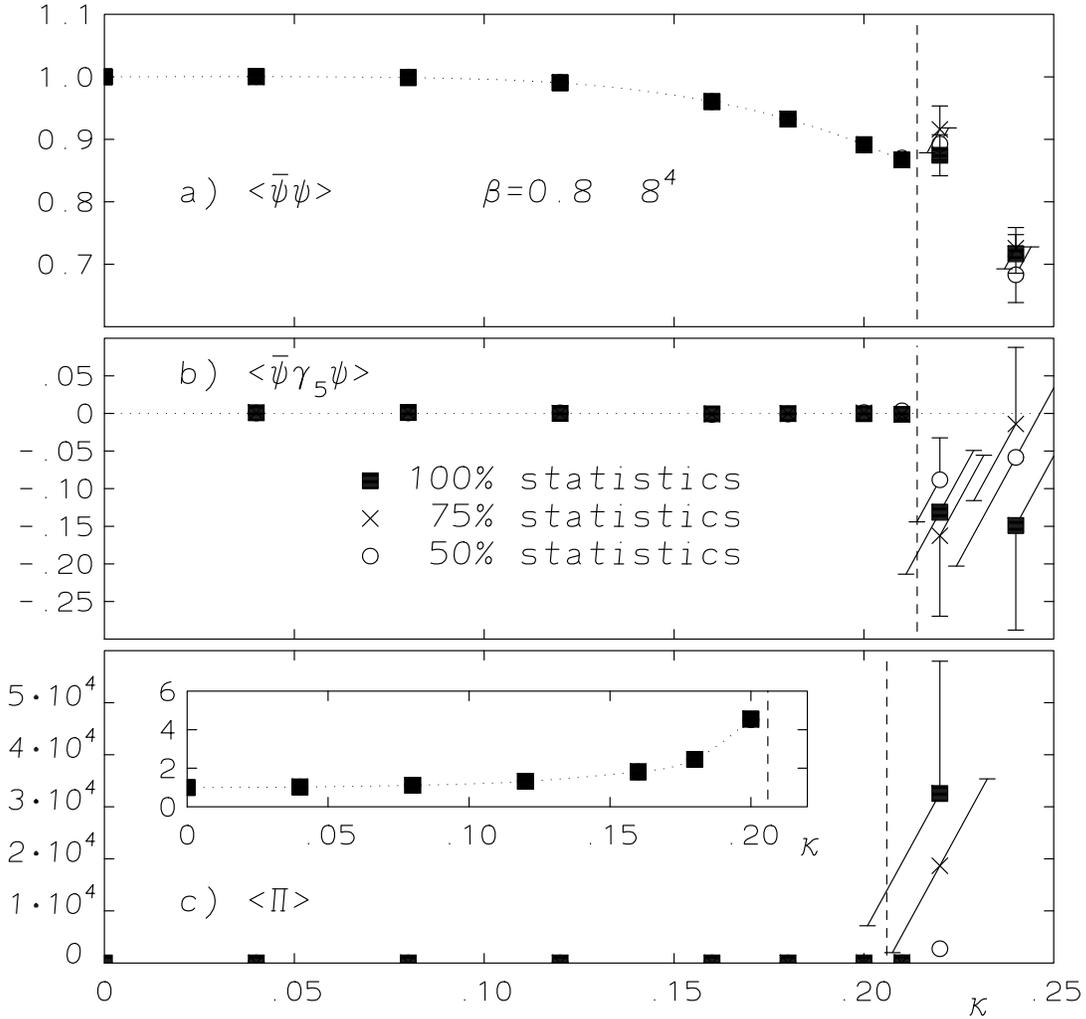}
     }
\end{center}
\vskip -5.5truecm
\caption{
The dependence on $~\kappa~$
 of $~\langle \psb \psi \rangle~$ ({\bf a}),
$~\langle \psb \gamma_5 \psi \rangle ~$ ({\bf b}) and 
$~\langle \Pi \rangle~$ ({\bf c}) at
$~\beta =0.8~$ on a $8^4$ lattice. Dotted lines have been added 
to guide the eye.
 }
\label{fig:bulk}
\vskip -0.2truecm
\end{figure}

To illustrate the problem we show in Figs.1a-c
the dependence of the 
observables $~\langle \psbps \rangle~$, $~\langle \psbgps \rangle~$ and
$~\langle\Pi\rangle~$ on $~\kappa~$ at $~\beta=0.8~$ on a $~8^4~$
lattice.
Well below the transition point, i.e., when
$~\kappa < \kappa_c^{\prime}(\beta ;N_{\tau};N_s) ~$ ,
these averages are statistically well-defined.
The increase of the number of measurements from, say, $~n = 100~$ 
to $~n =200~$ produces just a slight change of the averages,
and statistical errors decrease as $~\sim 1/\sqrt{n}~$. 
The situation however changes significantly
at $~\kappa \ageq \kappa_c^{\prime}(\beta ;N_{\tau};N_s)~$
(to the right of the vertical dashed line in Figs.1a-c).
The averages begin to fluctuate drastically with increasing $~n~$
(compare circles, crosses and squares in Figs.1a-c), and
the errorbars become dramatically large.
It is worthwhile to note
that the increase of statistics does {\it not} necessarily entail
the diminishing of the errorbars (compare, e.g., circles and
squares in Fig.1c). 
This means in fact that these averages and errors, which 
were calculated using the 
jackknife procedure, make {\it not} very much sense 
and the accumulation of measurements will not change the state of
affairs.
Note, that $~\kappa_c^{\prime}(\beta ;N_{\tau};N_s)~$ in 
Fig.1c is 
a little bit smaller than in Figs.1a,b, since 
$~\langle \Pi \rangle~$ is the more sensitive observable with respect to
small eigenvalues of 
$~\gamma_5 \cal{M}~$ (and $~\cal{M}~$ respectively).

The behaviour of $~\langle \psbgps \rangle~$ in Fig.1b
deserves maybe 
some additional comment. It was suggested \cite{aoki} that
at $~\kappa = \kappa_c(\beta )~$ there is a transition  to a 
parity--violating phase, in which $~\langle \psbgps \rangle \ne 0~$
(see also \cite{nasi}). 
At first sight our data for $~\langle \psbgps \rangle~$ in 
Fig.1b
look like the confirmation of this hypothesis, but in fact they are
{\it not}, because 
all averages and statistical errors to the right of the 
vertical dashed line 
are unreliable, and no definite conclusion can be drawn.

%
%
\begin{figure}[htb]
\begin{center}
\vskip -2.5truecm
\leavevmode
\hbox{
\epsfysize=570pt\epsfbox{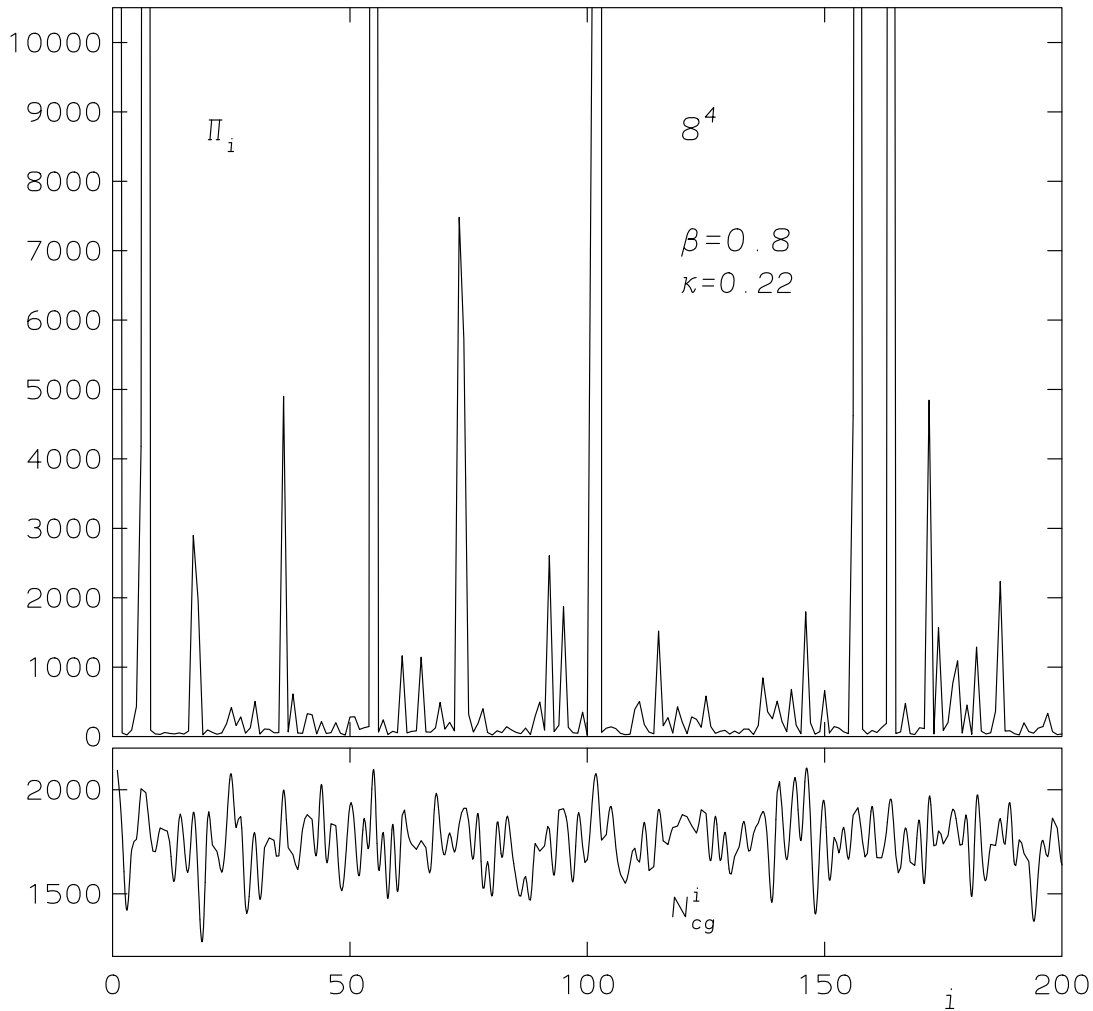}
     }
\end{center}
\vskip -5.5truecm
\caption{
The time history of $~\Pi~$ and $~N_{cg}~$ at
$~\beta =0.8~$ and $~\kappa =0.22~$ on a $8^4$--lattice.
 }
\label{fig:ncg}
\vskip -0.2truecm
\end{figure}
%
%
It is interesting to mention that on 'exceptional' configurations 
the number of conjugate gradient iterations $~N_{cg}~$ behaves 
differently from
the 'non--exceptional' case where the convergence of the
CG method is mainly determined by the condition number
$~\xi \equiv \lambda_{max}/\lambda_{min}~$ of the to be inverted
matrix \footnote{here $~\lambda_{max/min}~$ denotes
the maximal/minimal eigenvalue of the matrix to be inverted}.
 Moreover, when $~\lambda_{min}~$ is sufficiently small then 
$~N_{cg}~$ behaves as 
\eq
N_{cg} \sim \lambda_{min}^{-\frac{1}{2}}~
                                           \label{ncg1}
\en
\noi and therefore can be used as an indicator for approaching the chiral 
limit (see, e.g., \cite{blls,uka}). 
However, it is generally not known for which classes of distributions 
of eigenvalues eq.(\ref{ncg1}) holds.
In a special study of this problem using hermitian matrices with 
some predefined (but {\it continuous}) distributions 
of eigenvalues we could 
confirm the commonly accepted results \cite{qed2}.
Now, in case of 'exceptional' small eigenvalues the behaviour of
$~N_{cg}~$ does not follow eq.(\ref{ncg1}). 
This can be seen from  
			    Fig.\ref{fig:ncg}, where we display
the time histories of the pion norm $~\Pi~$ and $~N_{cg}~$ 
on a $~8^4~$ lattice slightly above $~\kappa_c(\beta)$.
While the pion norm $~\Pi~$ develops huge spikes (amplitudes up to
$~ \sim 10^5)~$ for some configurations, the values of 
$~N_{cg~}~$ vary within a $~\sim 10\%~$ corridor around 
the average value. 
Thus the CG convergence behaviour 
seems to be 
not only determined by $~\xi~$
but may also considerably depend on the eigenvalue distribution
of the matrix to be inverted.

\vspace{0.5cm}

%
\begin{figure}[htb]
\begin{center}
\vskip -2.5truecm
\leavevmode
\hbox{
\epsfysize=570pt\epsfbox{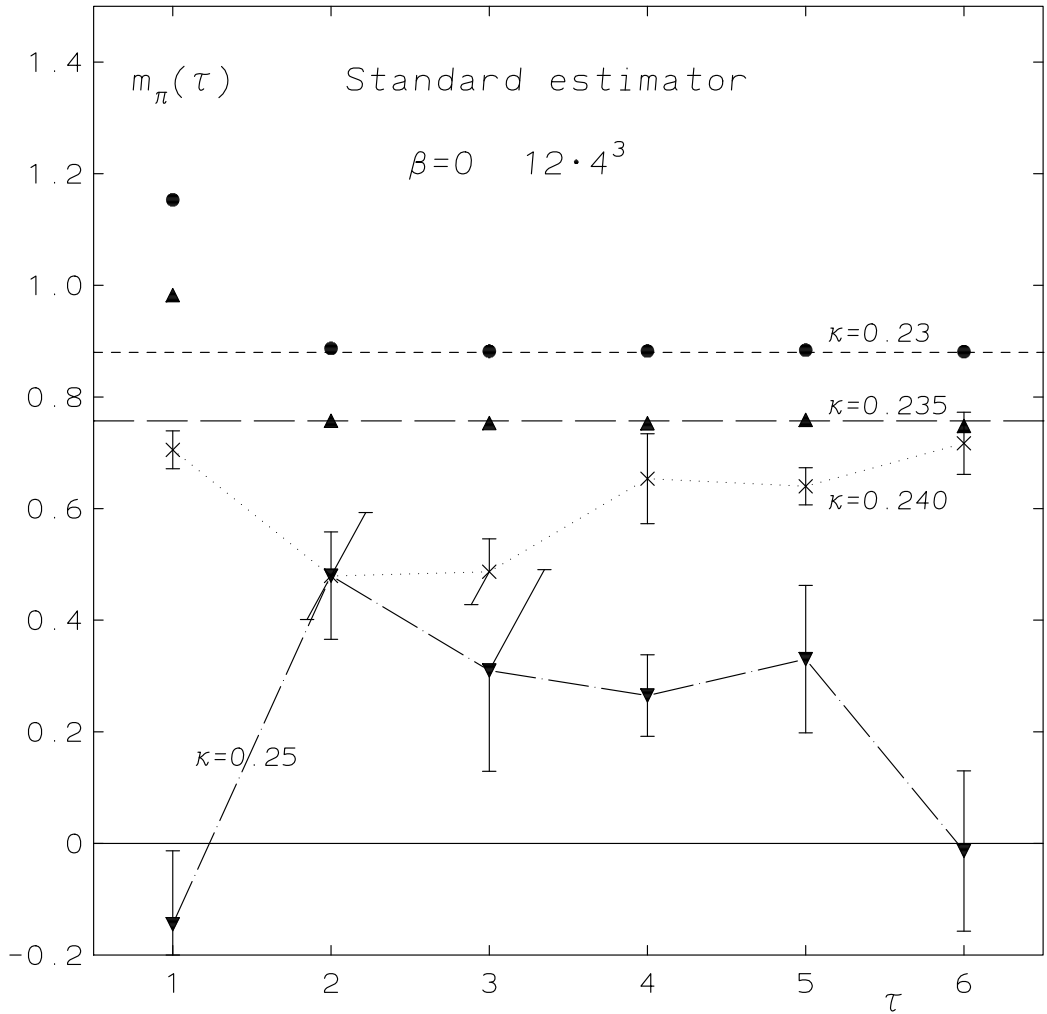}
     }
\end{center}
\vskip -5.5truecm
\caption{
The $~\tau$--dependence of the standard estimator
$~m_{\pi}(\tau)~$ 
at $~\beta=0$; the lattice 
size is $~12 \cdot 4^3$.
Lines are to guide the eye.
 }
\label{fig:stan}
\vskip -0.2truecm
\end{figure}

In Fig.3 we present
the $~\tau$--dependence of the standard
estimator $~m_{\pi}(\tau )~$ 
eq.(\ref{mass_stand}) for different $~\kappa$'s at 
$~\beta =0~$ on a $~12 \cdot 4^3~$ lattice. 
The value of $~\kappa_c~$ is expected to be $~\sim 1/4~$ at this
value of $~\beta~$ (at least  in the infinite volume).
For the $~12 \cdot 4^3~$ lattice the 'critical' region starts at
$~\kappa_c^{\prime}(\beta =0) \sim 0.24$. 
With increasing the lattice size this 'critical'
region shrinks : 
$~\kappa_c^{\prime}(\beta ; N_{\tau};N_s) \ra \kappa_c(\beta)~$
for $~N_{\tau},~N_s \ra \infty~$ \cite{bkr}.
Nevertheless, on a finite lattice the problem 
connected with the approach to the chiral limit
remains, and 
the usual 'safe' choice of $~\kappa~$ in QCD calculations, i.e., the 
choice of $~\kappa~$ well below $~\kappa_c(\beta )$, entails 
an unrealistically 
large
ratio $~m_{\pi}/m_{\rho}~$.

Sufficiently below the 'critical' value of $~\kappa~$ the 
standard estimator of the pion
mass eq.(\ref{mass_stand}) yields well--defined results.
The contributions of higher excited 
states die out with increasing $~\tau$, leaving a 
nice plateau in the $~\tau$--dependence of the effective mass
$~m_{\pi}(\tau )$, which
determines the actual mass at these values of couplings.
For the reasons given above, this estimator fails to give 
precise results in the region 
$~\kappa \ge \kappa_c^{\prime}(\beta ; N_{\tau};N_s)~$, i.e. when 
approaching the chiral limit.
(All data points of this figure represent averages of at least 5000
measurements).

\subsection{Another estimator for $~m_{\pi}~$ }

One possible way to bypass this problem is based on the observation
which we call the {\it factorization} of the contribution of 
near--to--zero eigenvalues.
In Figs.4a-b we display the values of
 $~\Gamma_i(\tau)~$ on individual
configurations $~i=1,\dots, 900~$ 
on a $~16\cdot 8^3~$ lattice at $~\beta=0~$ 
and $~\kappa=0.25~$ for $~\tau=3~$ (Fig.4a)
and $~\tau=4~$ (Fig.4b). 
The huge spikes (up to $~\sim 10^4$) in the time histories 
are due to the existence of very small eigenvalues in the 
spectrum of 
$~\gamma_5\cal{M}~$(respectively $~\cal{M}$).
These spikes appear with a certain frequency 
and just reflect the content of the spectrum of 
$~\gamma_5\cal{M}~$
for this value of $~\kappa$ which is close to $~\kappa_c(\beta)$
in a finite system, i.e. the lowest eigenvalues are allowed
to fluctuate from configuration to configuration.

The important observation is the absence of such peaks
in the ratio of the individual correlators 
\eq
g_i(\tau ) =
\frac{ \Gamma_i(\tau  +1) }{ \Gamma_i(\tau ) } 
                                                      \label{ratio}
\en
\noi which means the factorization of 
the 'divergent' contributions
originating from near--to--zero eigenvalues.
The statistical fluctuations of this ratio are very small
even at $~\kappa ~\ageq ~\kappa_c(\beta)~$ (see Fig.4c)
as compared with fluctuations of $~\Gamma_i(\tau )$
(Figs.4a,b).

Therefore, we propose to use the ratio eq.(\ref{ratio})
for the extraction of the 'pion' mass 
near the chiral transition. 
Then a new estimator for the pseudoscalar mass, which we 
denote by $~\mpr_{\pi}(\tau)$ , can be obtained from the 
following expression

\eq
\frac{\cosh \mpr_{\pi}(\tau ) \left( \frac{N_{\tau}}{2} - \tau -1 \right) }
{\cosh \mpr_{\pi}(\tau ) \left( \frac{N_{\tau}}{2} - \tau  \right) }
\equiv \frac{1}{n} \sum_{i=1}^{n} g_i (\tau )  ~.
                 \label{mass_new2}
\en

\noi Therefore, the relation to the standard estimator $~m_{\pi}(\tau)~$ in
eq.(\ref{mass_stand}) is given by

\eqa
\frac{\cosh m_{\pi}(\tau ) \left( \frac{N_{\tau}}{2} - \tau -1 \right) }
{\cosh m_{\pi}(\tau ) \left( \frac{N_{\tau}}{2} - \tau  \right) }
= \frac{\cosh \mpr_{\pi}(\tau ) \left( \frac{N_{\tau}}{2} - \tau -1 \right) }
{\cosh \mpr_{\pi}(\tau ) \left( \frac{N_{\tau}}{2} - \tau  \right) }
+\frac{1}{n} \sum_{i=1}^{n}  g_i (\tau )  \cdot
\frac{ \delta \Gamma_i(\tau ) }{ \hg(\tau ) }~,
                                            \label{mass_new1}
\ena

\noi with 
$$
\delta \Gamma_i(\tau ) = \Gamma_i(\tau ) - \hg(\tau )~;
\qquad
\sum_{i}^{n} \delta \Gamma_i(\tau ) \equiv 0 ~ .
$$

%
%
\begin{figure}[htb]
\begin{center}
\vskip -2.5truecm
\leavevmode
\hbox{
\epsfysize=570pt\epsfbox{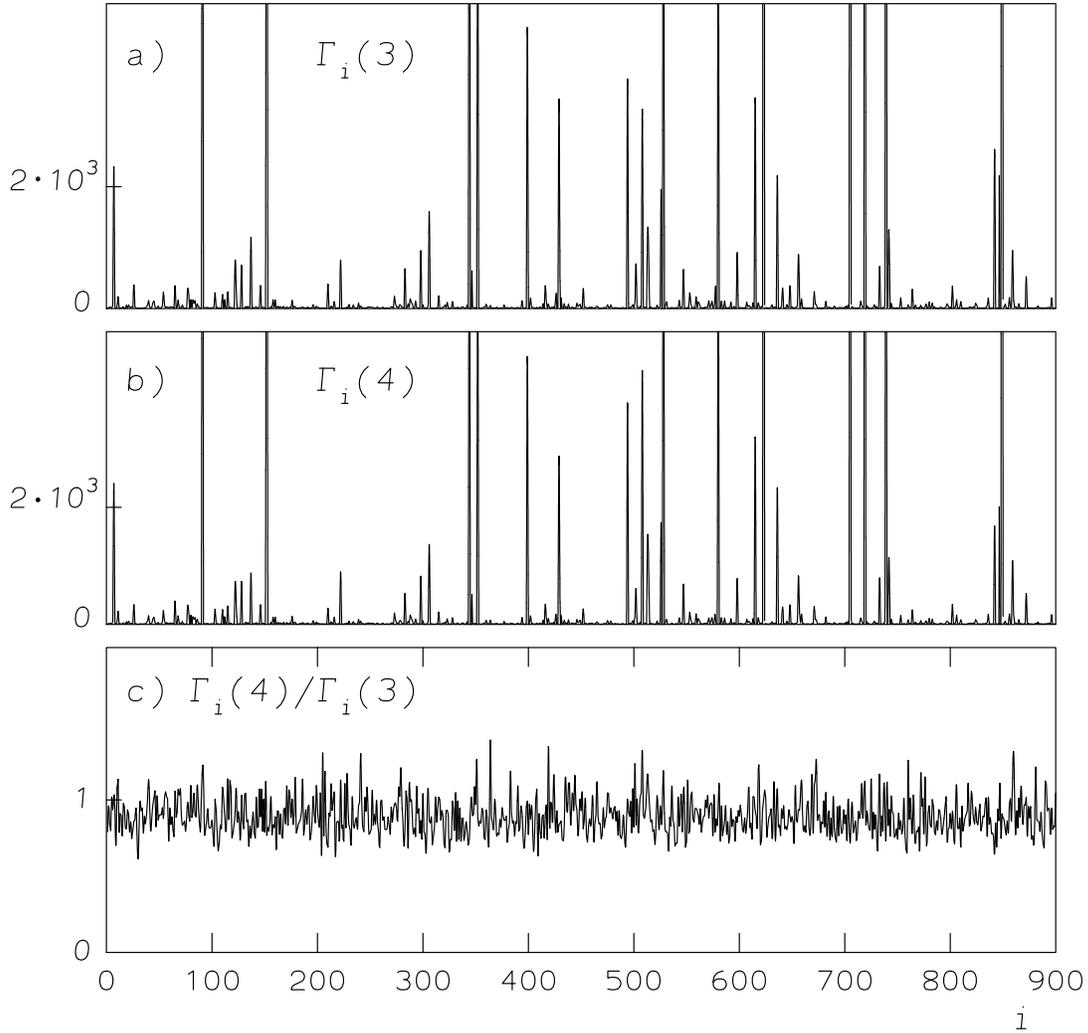}
     }
\end{center}
\vskip -5.3truecm
\caption{
Time histories $~\left\{ \Gamma_i(\tau) \right\}~$ for
$~\tau=3~$ ({\bf a}), $~\tau=4~$ ({\bf b}) and
$~\left\{ \Gamma_i(\tau=4)/\Gamma_i(\tau=3) \right\}~$ ({\bf c})
on a $~16 \cdot 8^3~$ lattice at $~\beta=0~$ and $~\kappa=0.25$.
 }
\label{fig:hist}
\vskip -0.2truecm
\end{figure}
%

%
%
\begin{figure}[tp]
\begin{center}
\vskip -2.5truecm
\leavevmode
\hbox{
\epsfysize=570pt\epsfbox{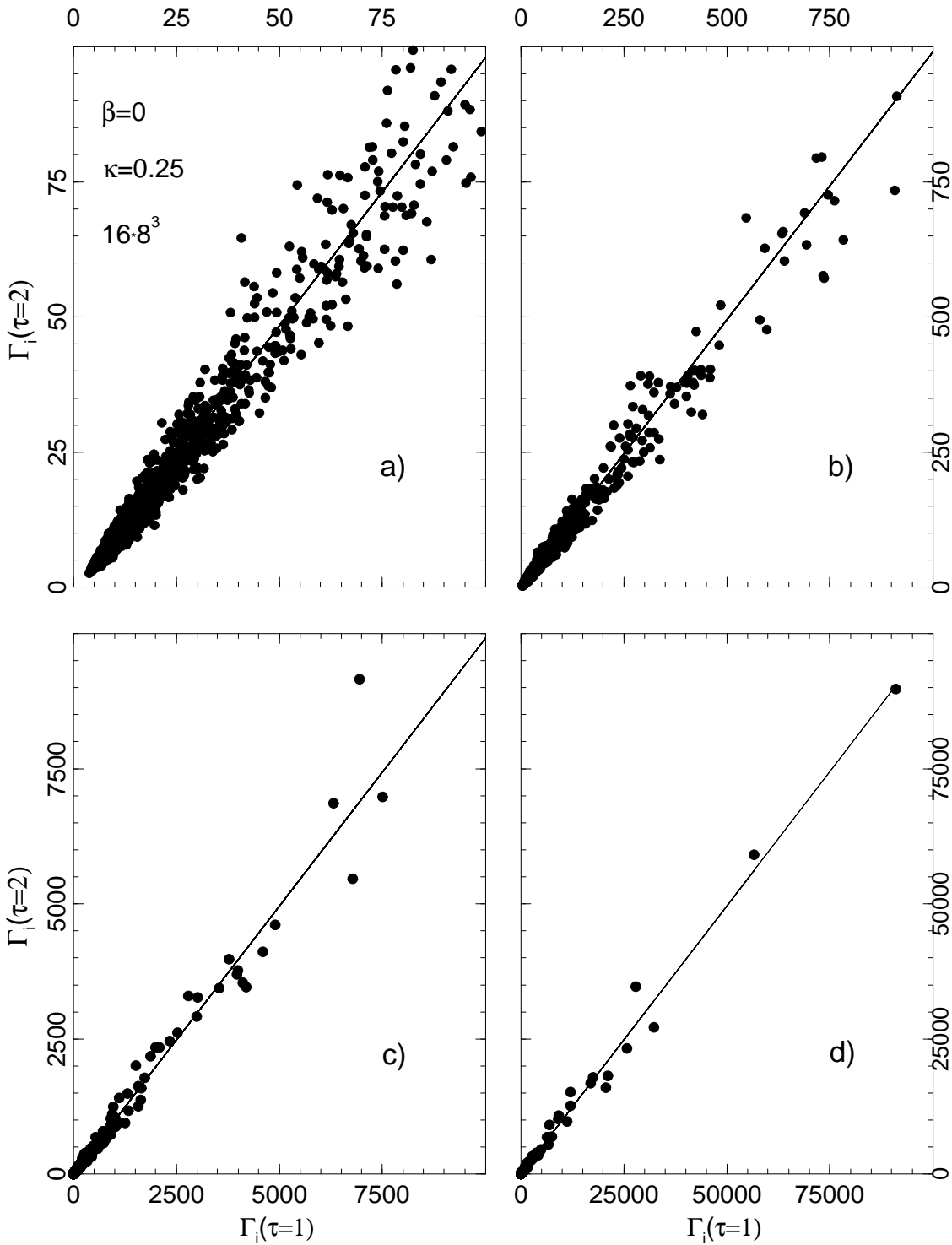}
     }
\end{center}
\caption{
$\Gamma_i(\tau =2)$ as a function of $\Gamma_i(\tau =1)$
at $\beta =0$ and $\kappa =0.25$ on a $16 \cdot 8^3$ lattice 
at different scales.  Single points represent individual
configurations.
Straight lines correspond to the linear fit according to
the least squares method.
 }
\label{fig:4_prim}
\vskip -0.2truecm
\end{figure}

\noi In general, of course, this is not the correct way 
to calculate masses. 
However, under certain circumstances it is justified to do so.

Assume that $~x~$ and $~y~$ are two {\it correlated}
random variables with some distribution $~P(x;y)$,
where $~\int \!\! dxdy \, P(x;y)=1$.
Then the average of any functional $~{\cal O}(x;y)~$ is

\eq
\langle {\cal O} \rangle = \int \!\! dxdy \, 
{\cal O} \cdot P(x;y)~.
\en

\noi One can define the conditional average $~{\bar y}(x)~$ as

\eq
{\bar y}(x) = \int \!\! dy \, y P(x;y) \Big/
\!\! \int \!\! dy \, P(x;y)~.
\en

\noi If the distribution $~P(x;y)~$ is such that

\eq
{\bar y}(x) = C \cdot x~,\quad C=const~,
             \label{cond_aver}
\en
\noi then it is easy to see that

\eq
\frac{\langle y \rangle }{\langle x \rangle }
= \left\langle \frac{y}{x} \right\rangle.
\en

\noi In our case $~x = \Gamma_i(\tau )~$ and
$~y = \Gamma_i(\tau +1 )$. 
In principle, one can use also the ratios 
$~\Gamma_i(\tau+k)/\Gamma_i(\tau)~$ with
$~k \ge 2$, though the  
signal--to--noise ratio decreases with increasing $~k$.

In Fig.\ref{fig:4_prim} we show 
$~\Gamma_i(2 )~$ as a function of $~\Gamma_i(1)~$ 
at $~\beta =0~$ and $~\kappa =0.25~$ on a $~16 \cdot 8^3~$
lattice.
For illustrative purposes we have shown our data
of $\sim 2000~$ measurements with four different 
choices of the scale. 
The distribution gives a clear indication in favour of
eq.(\ref{cond_aver}).
The correlation coefficient $~\rho~$ defined in a standard
way \cite{vdv} is very close to unity, and varies between
$~\sim 0.98~$ and $~\sim 0.995~$ for different subsets 
of measurements shown in 
Fig.\ref{fig:4_prim}a $~\div ~$ Fig.\ref{fig:4_prim}d.

To provide a proof that $~\mpr_{\pi}(\tau)~$
can serve as a reliable estimator of the pion mass 
$~m_{\pi}(\tau)~$ we will compare the properties of both
estimators.
We'll show that for $~\kappa$--values sufficiently below 
$~\kappa_c(\beta)~$, i.e., where the
standard estimator of $~m_{\pi}~$ can be reliably defined, both
estimators are in a very good agreement. 
For values of $~\kappa~$ very close to $~\kappa_c(\beta)$, where
the standard estimator fails to work, while 
$~\mpr^2_{\pi}~$ fits {\it the same straight line}.

We have run simulations in the confinement phase at $~\beta = 0$ and
$~\beta = 0.8~$ using lattices with $~N_{\tau} = 12;~16;~20~$ 
and $~N_s =4;~8~$.
For the matrix inversion we used the standard
conjugate gradient method \cite{cg1,cg2}
with even--odd decomposition  \cite{deg1,deg2}
which guaranties convergence even at
$~\kappa ~\ageq ~ \kappa_c(\beta)$. \\
%
%
%
\begin{figure}[htb]
\begin{center}
\vskip -2.5truecm
\leavevmode
\hbox{
\epsfysize=570pt\epsfbox{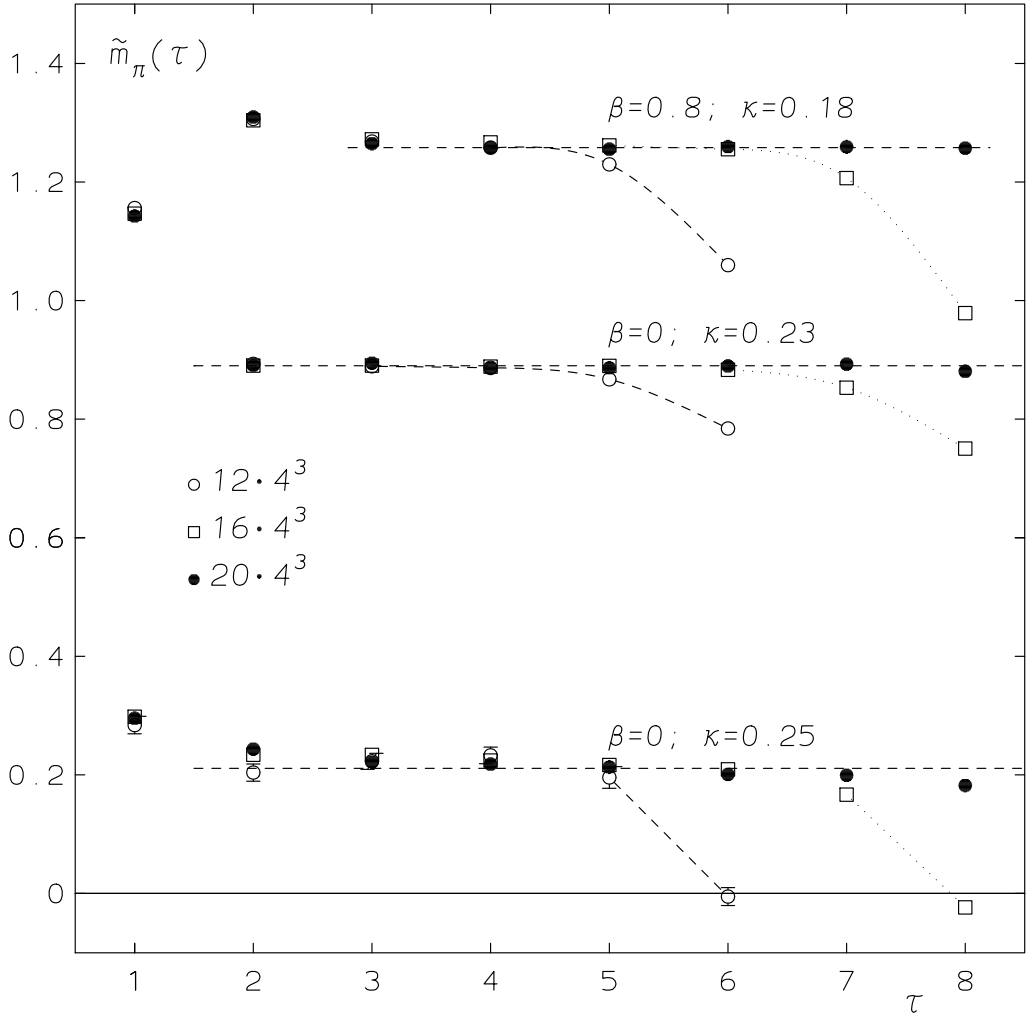}
     }
\end{center}
\vskip -5.5truecm
\caption{
The $~\tau$--dependence of 
$~\mpr_{\pi}(\tau)~$ at $~\beta=0~$ and $~\beta=0.8~$.
Lattice sizes are $~12 \cdot 4^3~$, $~16 \cdot 4^3~$ and $~20 \cdot 4^3$.
Lines are to guide the eye.
 }
\label{fig:time}
\vskip -0.2truecm
\end{figure}
The $~\tau$--dependence of the effective
mass $~\mpr_{\pi}(\tau )~$
appears to be more complicated than that of the standard estimator
$~m_{\pi}(\tau )~$.
In Fig.6 we have plotted
the dependence of $~\mpr_{\pi}(\tau )~$  on
$~\tau~$ at different $~\beta$'s and $~\kappa$'s on lattices
with $~N_s =4~$ and $~N_{\tau} = 12;~16; ~20~$.  At $~\beta =0~$
the value $~\kappa =0.23~$ is sufficiently below the transition point 
$~\kappa_c(0) \sim 0.25~$ 
(no appearance of 'exceptional' small eigenvalues 
of $~\cal{M}~$ and $~\gamma_5 \cal{M}~$ respectively)
and therefore $~\mpr_{\pi}(\tau )~$ can be compared with 
the corresponding $~m_{\pi}(\tau )~$ shown in Fig.3. 
The standard estimator $~m_{\pi}(\tau )~$ exhibits a nice plateau up to
$~\tau =6~$ on a $~12 \cdot 4^3~$ lattice, while the corresponding
plateau for $~\mpr_{\pi}(\tau )~$ 
is to be seen only for $~\tau \aleq 5~$.
In this case both estimators ($m_{\pi}(\tau )~$ and
$~\mpr_{\pi}(\tau )~$) are in a very good agreement 
as long as $~\tau~$ is not too close to $N_{\tau}/2$.
These deviations from the standard $m_{\pi}(\tau)$
at $~\tau \sim N_{\tau}/2~$ are a general feature of the new estimator 
which presumably stem from the way of averaging the corresponding
observables on a finite lattice 
(but which fade with increasing $~N_s$, see below).
The increasing of $~N_{\tau}~$  entails the 
extension of the plateau.  The same effect, i.e. the enlarging of the 
plateau in the $~\tau$--dependence
of  $~\mpr_{\pi}(\tau )~$ occurs for other $~\beta$'s
and $~\kappa$'s, including values $~\kappa ~\ageq ~\kappa_c(\beta)~$
where the standard estimator is not well--defined 
(e.g., $~\kappa =0.25; ~\beta =0~$ in Fig.6~).
It is important to notice that increasing $~N_{\tau}~$ does
{\it not} change the position of the plateau, influencing only its
extension.

%
%
\begin{figure}[htb]
\begin{center}
\vskip -2.5truecm
\leavevmode
\hbox{
\epsfysize=570pt\epsfbox{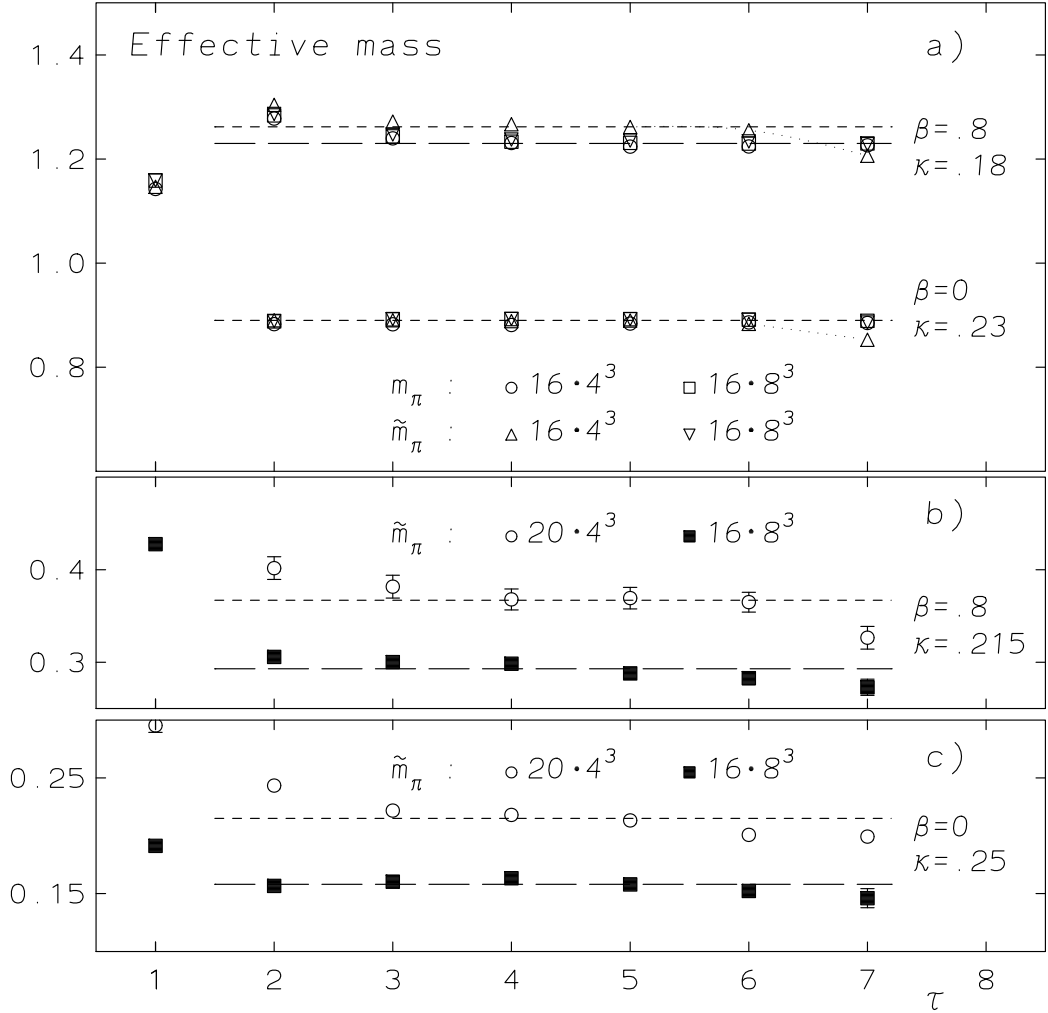}
     }
\end{center}
\vskip -5.5truecm
\caption{
The $~\tau$--dependence of
$~m_{\pi}(\tau)~$ and $~\mpr_{\pi}(\tau)~$
at $~\beta=0~$ and $~\beta=0.8~$ for two lattices :
$~16 \cdot 4^3~$ and $~16 \cdot 8^3~$ ({\bf a});
the $~\tau$--dependence of $~\mpr_{\pi}(\tau)~$
at $~\beta=0.8~$ ({\bf b}) and $~\beta=0~$ ({\bf c}) for lattices 
$~20 \cdot 4^3~$ and $~16 \cdot 8^3$. Lines are to guide the eye.
 }
\label{fig:space}
\vskip -0.2truecm
\end{figure}

Finite volume effects, i.e., the dependence of $~\mpr_{\pi}(\tau )~$ on
$~N_s~$ tend to increase slightly with increasing coupling $~\beta$
 -- this could be interpreted as an extension of the 'critical' zone 
in $~\kappa~$ with rising $~\beta~$.
In Fig.7a we compare
the $~\tau$--dependence of both estimators
$~m_{\pi}(\tau )~$ and $~\mpr_{\pi}(\tau )~$ at $~\beta =0~$ and $~\beta
=0.8~$ for two lattice sizes :  $~16 \cdot 4^3~$ and $~16 \cdot 8^3~$.
The values of 
the corresponding $~\kappa$'s are chosen to be
sufficiently far from the transition point, ensuring 
the applicability of the standard definition of the pion mass.
At $~\beta =0~$ both estimators are in an excellent agreement even on the
smaller lattice with $~N_s =4~$.  At $~\beta =0.8~$ 
$~\mpr_{\pi}(\tau )~$
on the smaller lattice gives an $\approx$3\% overestimated value.
However, on the lattice with $~N_s =8~$ the agreement between both
estimators becomes very good.
Note, that at these $~\kappa$--values the deviations of 
$~\mpr_{\pi}(\tau )~$ from 
the plateau for $~\tau \rightarrow N_{\tau}/2~$ as described in 
Fig.6 decreased substantially.

In Fig.7b and Fig.7c we 
compare the values of 
$~\mpr_{\pi}(\tau )~$ at the same $~\beta~$'s as in Fig.7a
but now at $~\kappa$'s chosen near $~\kappa_c(\beta)~$.
The finite volume dependence becomes stronger when $~\kappa~$ becomes
close to $~\kappa_c(\beta)$. 
The estimate of $~\mpr_{\pi}(\tau )~$
is lowered by increasing the spatial extension $~N_s~$.
The quality of the plateau is good enough to determine 
$~\mpr_{\pi}(\tau )~$ with good accuracy.

%
%
\begin{figure}[htb]
\begin{center}
\vskip -2.5truecm
\leavevmode
\hbox{
\epsfysize=570pt\epsfbox{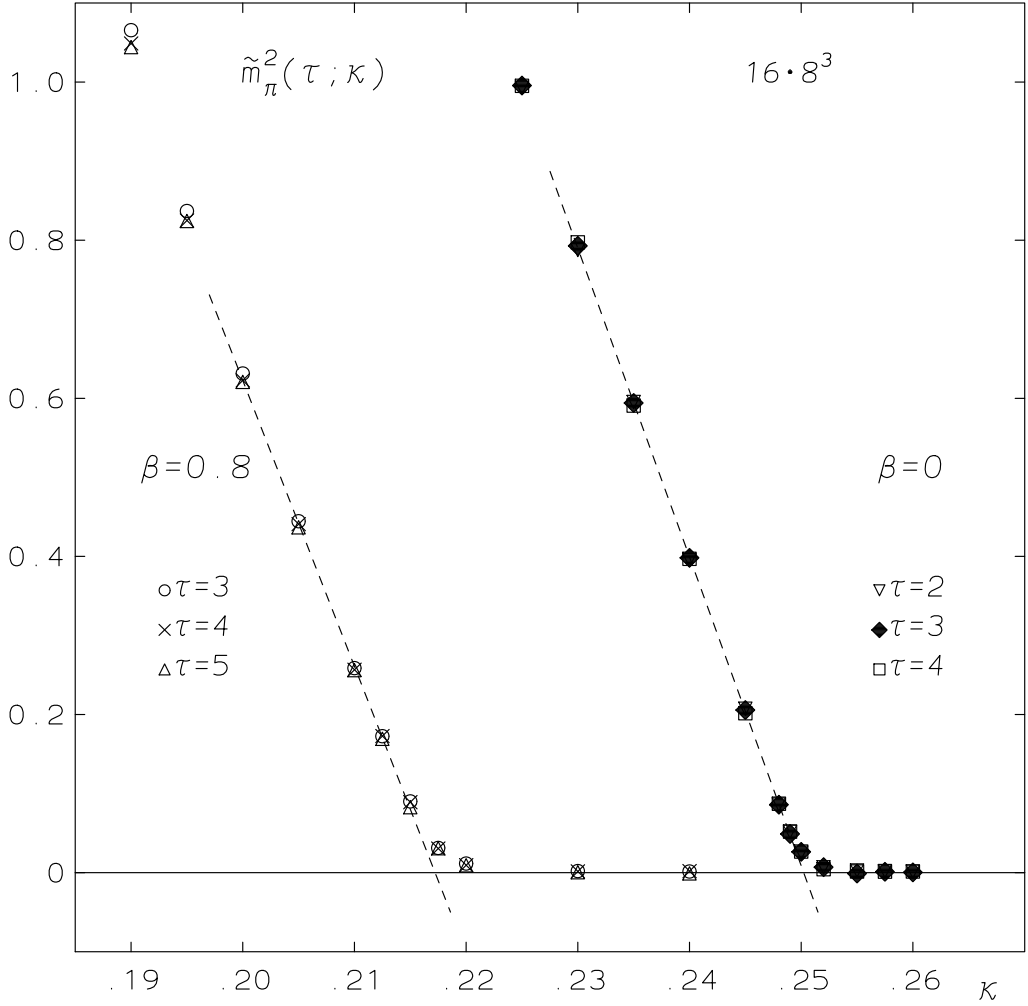}
     }
\end{center}
\vskip -5.5truecm
\caption{
The $~\kappa$--dependence of $~\mpr_{\pi}^2(\kappa)~$ at $~\beta=0~$ and
$~\beta=0.8~$ on a $~16 \cdot 8^3~$ lattice for several values
of $~\tau$. The errorbars are smaller than the symbol size.
Lines drawn are to guide the eye.
 }
\label{fig:pmass_n}
\vskip -0.2truecm
\end{figure}

\vspace{0.25cm}
In Fig.8 we show the $~\kappa$--dependence of
$~\mpr^2_{\pi}~$ for $~\beta =0~$ and
$~\beta=0.8$ on a $~16\cdot 8^3$ lattice. 
Different symbols correspond to different values of $~\tau$. 
At both values of $~\beta~$ the 
new estimator $~\mpr^2_{\pi}~$
behaves in a similar way.  Even at $~\kappa ~\ageq ~ \kappa_c(\beta)~$
the values of $~\mpr_{\pi}^2~$ at different $~\tau~$ 
and $~\beta~$ are in a good agreement.
For both $~\beta$--values $~\mpr^2_{\pi}~$ shows a linear dependence
on $~\kappa~$ in the vicinity of $~\kappa_c(\beta)$.
This leads to the  behaviour

\eq
\mpr^2_{\pi} = B_{\pi}(\beta) \cdot m_q~; \qquad m_q \ra 0~,
                                          \label{b}
\en

\noi where the dimensionless bare fermion mass $~m_q~$ is defined as

\eq
m_q = \frac{1}{2\kappa} - \frac{1}{2\kappa_c(\beta)} ~.
\nonumber
\en

\noi At $~\beta=0~$ the  straight line extrapolation (broken line in 
Fig.8) predicts for the 'critical' value
$~\kappa_c(0) \simeq 0.2502(1)~$
which is in a very good agreement with the strong coupling results \cite{kawa}.
With increasing $~\beta~$ the value $~\kappa_c(\beta )~$ decreases 
: $~\kappa_c(\beta =0.8) \simeq 0.2171(1)~$ (errorbars correspond 
to one standard deviation in $~\chi^2/n_{d.o.f.}~$).
The corresponding slopes are : $~B_{\pi}(0) = 4.91(4)~$
and $~B_{\pi}(0.8) = 3.42(3)$.

For $~\kappa$--values sufficiently below $~\kappa_c(\beta)~$, i.e.,
for $~\kappa < \kappa_c^{\prime} (\beta)~$,
$~\mpr^2_{\pi}~$ is just the same as the standard
estimator $~m^2_{\pi}~$. For $~\kappa$'s very close to
$~\kappa_c(\beta)~$ the standard estimator 
can't be applied, still
$~\mpr^2_{\pi}~$ fits very well the same straight
line having a very small statistical error.
Our data at $~\beta=0~$ suggest, that for 
$~\kappa ~\ageq ~0.246 ~ ( m_q ~ \aleq ~ 0.034)~$
the standard estimator of $~m_{\pi}^2~$ deviates from the 
straight--line behaviour in Fig.8.
For instance, at $~ \kappa \ge 0.248~(  m_q ~ \aleq ~0.0176)~$
the standard estimator was not applicable any more.
Numbers change when considering the case $~\beta=0.8$. 
Here the deviation of $~m_{\pi}^2~$ from the straight--line behaviour starts 
already at $~m_q \simeq 0.078$, which corresponds to $~\kappa =0.21~$.
This demonstrates the advantage of $~\mpr_{\pi}~$
which gives the possibility to approach the chiral limit much closer 
than the standard estimator of $~m_{\pi}~$ would allow.
Especially, systematic errors induced by the finite volume can be investigated 
in the 'critical' region by means of $~\mpr_{\pi}~$ with good 
accuracy, which in turn should allow a more precise extrapolation to 
the thermodynamic limit. 

\section{The subtracted chiral condensate}

One can use the Goldstone theorem to define the 'physical'
(subtracted) chiral condensate $~\langle \psb \psi \rangle_{subt}~$.
Following \cite{maian} we define 
$~\langle \psb \psi \rangle_{subt}~$ which is
supposed to serve as an order parameter for the chiral symmetry
breaking (at least, in the continuum limit)

\eqa
\langle \psb \psi \rangle_{subt} &=& \lim_{\kappa \ra \kappa_c}
\langle \psb \psi \rangle_{subt} (\kappa)~;
\nonumber \\
\nonumber \\
\langle \psb \psi \rangle_{subt} (\kappa) &=&
(\frac{1}{\kappa} - \frac{1}{\kappa_c} )
\cdot Z_P \sum_x \langle \psb \gamma_5 \psi_x 
\cdot \psb \gamma_5 \psi_0 \rangle~,
                  \label{cond}
\ena

\noi where $~Z_P~$ is the renormalization constant of the pseudoscalar
quark density and the sum in the r.h.s. of eq.(\ref{cond}) is 
connected with the pion norm $~\langle \Pi \rangle~$.
At nonzero lattice spacing the pion mode in the limit 
$~\kappa \ra \kappa_c(\beta )~$ should result in a constant contribution
to the condensate  $~\langle \psb \psi \rangle_{subt} ~$
while the contribution of the higher energy (massive) modes should
decrease as $~\sim (\kappa_c -\kappa )~$.

%
%
\begin{figure}[htb]
\begin{center}
\vskip -2.5truecm
\leavevmode
\hbox{
\epsfysize=570pt\epsfbox{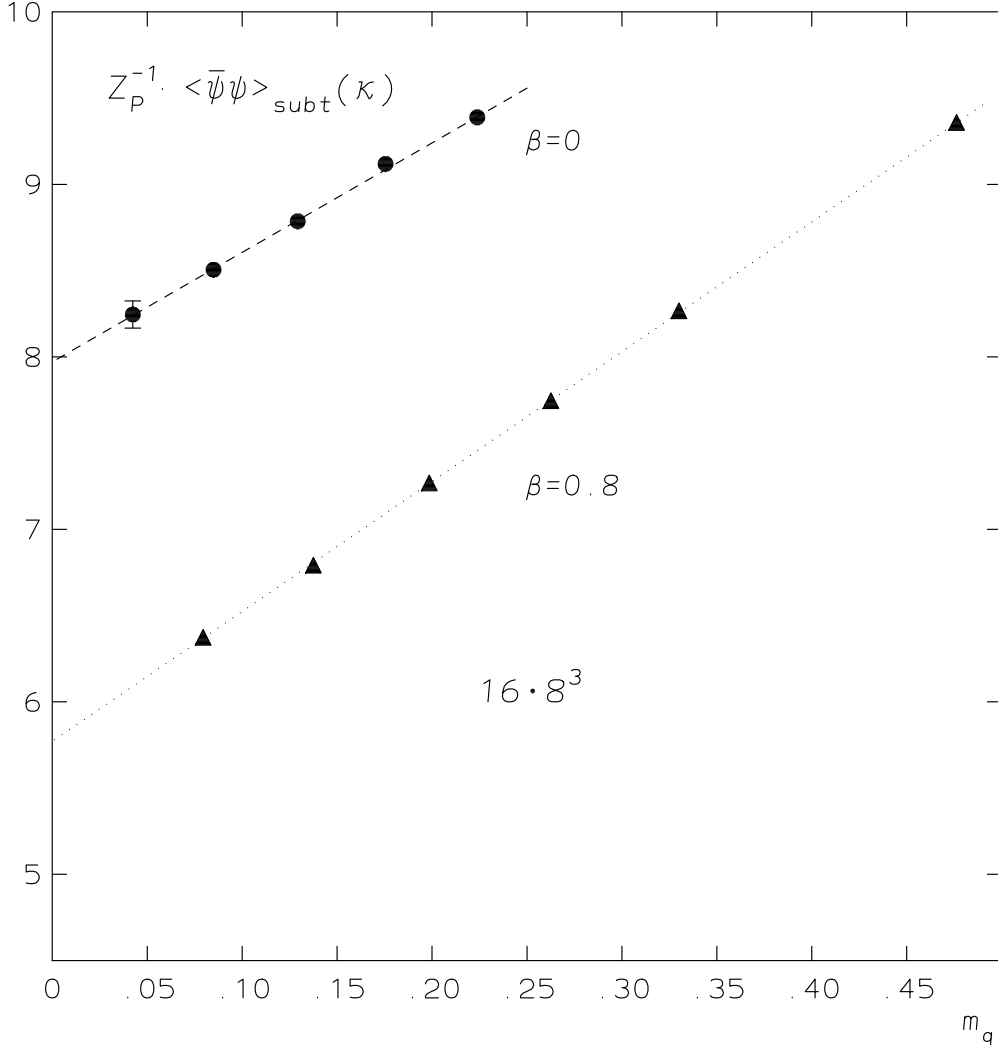}
     }
\end{center}
\vskip -5.5truecm
\caption{
The $~m_q$--dependence of the subtracted condensate
$~Z_P^{-1} \cdot \langle \psb \psi \rangle_{subt}(\kappa)~$ at
$~\beta =0~$ and $~\beta =0.8~$ on a $~16 \cdot 8^3$--lattice.
Lines are to guide the eye.
 }
\label{fig:cond}
\vskip -0.2truecm
\end{figure}

Fig.9 shows
the dependence of
$~\langle \psb \psi \rangle_{subt}(\kappa) \cdot Z_P^{-1}~$  on $~\kappa~$
for $~\beta = 0~$ and $~\beta = 0.8~$. Here we have chosen $~\kappa~$
values in the region where near--to--zero 'exceptional' modes
not yet show up. 
The pion mass $~m_{\pi}~$ reached already the asymptotic 
regime shown in Fig.8.
For these $~\kappa$'s the values of the condensate
are on the (expected) straight line which
permits to make a reasonable extrapolation to 
$~\kappa \ra \kappa_c(\beta)~$.
The extrapolated values of $~\langle \psb \psi \rangle_{subt} (\kappa)~$ 
are nonzero, of course (at least, because of the finite spacing).

The main observation here is that the values of the condensate 
are comparatively large,
comparing with what one could expect from the naive definition
$~~\langle \psb \psi \rangle_{bare} - \langle \psb \psi \rangle_{pert}~~$
with $~~\langle \psb \psi \rangle_{bare} \leq 1~$
(although one cannot exclude that the renormalization constant $~Z_P~$ is
responsible for this effect).

A possible interpretation of this effect of {\it amplification} of
$~\langle \psb \psi \rangle_{subt} \cdot Z_P^{-1}~$
is due to the unphysical contributions in the quenched approximation.
In the dynamical fermion case the fermionic determinant tends to
decrease the pion norm substantially comparing with that in the
quenched approximation \cite{qed2} (at least in the confinement phase).
Therefore, we expect that in the
dynamical fermion case the condensate 
should be much smaller. From another point of 
view the unphysical contributions in
the quenched approximation were discussed  also in \cite{log1,log2,log3}.

\vspace{0.2cm}
At $~\kappa \sim \kappa_c(\beta)~$ the behaviour of 
$~\langle \psb \psi \rangle_{subt}(\kappa)~$ should be mainly
controlled by the value of 
the preexponential factor $~A_{\pi}~$ in the pseudoscalar correlator
$~\Gamma (\tau )~$ in eq.(\ref{transf}). 
The existence of the pion requires $~A_{\pi}~$ to behave as
$ A_{\pi} \sim 1/m_{\pi}~$ when $~ m_{\pi} \ra 0~$.
Together with eq.(\ref{b}) this implies

\eqa
A_{\pi} \sim \frac{1}{\sqrt{m_q}}~, \qquad m_q \ra 0.
					 \label{a2}
\ena

\noi In Fig.10 we present the value $~A_{\pi}^2 \cdot m_q~$
as a function of $~m_q~$ obtained from our
data on a $~16 \cdot 8^3~$ lattice.
For bare masses $~m_q < 0.06~$ we are not able
to indicate $~A_{\pi}^2\cdot m_q$, since the extracted 
$~A_{\pi}$'s become statistically unreliable. 
At $~\beta=0~$ the value $~A_{\pi}^2 \cdot m_q~$
shows a gradual increase with decreasing $~m_q$, which, however, is
not in contradiction to eq.(\ref{a2}), taking into account
the comparatively large values of $~m_q$.
Remarkably, we observe $~A_{\pi}^2 \cdot m_q~$ to fit a constant 
value at $~\beta=0.8~$ with good accuracy in agreement 
with eq.(\ref{a2}). 

%
%
\begin{figure}[htb]
\begin{center}
\vskip -2.5truecm
\leavevmode
\hbox{
\epsfysize=570pt\epsfbox{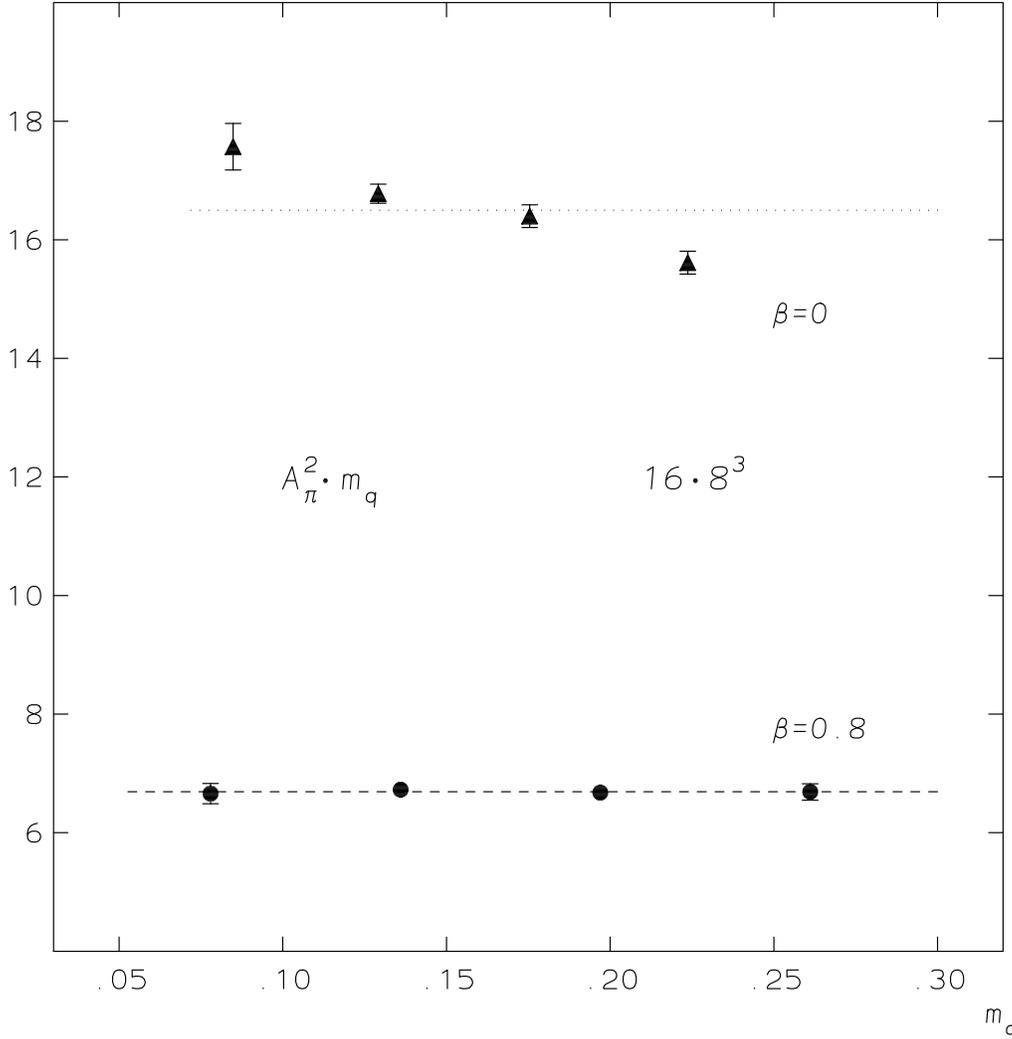}
     }
\end{center}
\vskip -5.5truecm
\caption{
$~A_{\pi}^2 \cdot m_q~$ vs. $~m_q~$ at $~\beta=0~$ and $~\beta=0.8~$
on a $~16\cdot 8^3~$  lattice.
Added lines are to guide the eye.
 }
\label{fig:aa_mq}
\vskip -0.2truecm
\end{figure}

\section{Conclusions}

In this work we have studied the chiral limit 
of a lattice gauge theory with Wilson fermions within the quenched
approximation, employing compact lattice QED in the confinement phase.

We observed that ratios of the pseudoscalar correlators
$~\Gamma_i(\tau +1)/\Gamma_i(\tau)~$ do {\it not} suffer
from near--to--zero ('exceptional') eigenmodes of the fermionic matrix.
This means that for every given configuration all 'divergent'
contributions to the correlators are {\it factorized}.

Making use of this observation 
we propose another estimator  $~\mpr_{\pi}~$
of the pseudoscalar mass, which is well defined 
near the chiral transition line $~\kappa_c(\beta)~$ and even for
$~\kappa > ~\kappa_c(\beta )~$
in contrast to the standard estimator $~m_{\pi}$.

It should be stressed that the possibility to introduce 
this estimator is based on the special kind of
correlations between $~\Gamma_i(\tau )~$ with different
$~\tau ~$, i.e., such that eq.(\ref{cond_aver}) is
fulfilled. This is not a universal property, of course.

For $~\kappa$--values sufficiently below $~\kappa_c(\beta)~$, i.e., where the
standard estimator of $~m_{\pi}~$ can be reliably defined, both
estimators are proved to be in a very good agreement. 
By approaching $~\kappa_c(\beta)~$ we observe a linear dependence of
$~m_{\pi}^2~$ on $~\kappa~$:
$$~m^2_{\pi} \sim \Bigl (~\kappa - \kappa_c(\beta) ~\Bigr )~. ~$$ 
For values of $~\kappa~$ very close to $~\kappa_c(\beta)~$
the standard estimator fails to work, while 
$~\mpr^2_{\pi}~$ still fits the same straight line
having a very small statistical error.

This study leads us to the conclusion that the main contribution to the
pion mass comes from larger modes, and the 'exceptional'
near--to--zero eigenvalues
of the fermionic matrix $~\gamma_5 \cal{M}~$ (and $\cal{M}~$respectively) 
play {\it no} physical role.  This conclusion is in agreement with
\cite{lue1}.

The new estimator $~\mpr_{\pi}~$
gives the possibility to approach the chiral limit much closer 
than the standard estimator of $~m_{\pi}~$ would allow.
Especially, systematic errors induced by the finite volume can be investigated 
in the 'critical' region by means of $~\mpr_{\pi}~$ with good 
accuracy, which in turn should allow a more precise extrapolation to 
the thermodynamic limit. 

The 'critical' value $~\kappa_c(\beta)~$ can be determined
with high accuracy. Our $~\kappa_c(0)~$ is in very good 
agreement with strong coupling predictions \cite{kawa}.

We investigated the subtracted chiral condensate
$~~\langle \psb \psi \rangle_{subt}~~$ 
following the definition in \cite{maian}
near the chiral transition line $~\kappa_c(\beta)$.
We obtained a comparatively large value of 
$~~\langle \psb \psi \rangle_{subt} \cdot Z_P^{-1}~~$ which we interpret as a
possible effect of the quenched approximation.

It would be very interesting to know in which way the fermionic sea 
--dynamical fermions-- could influence the behaviour of 
the pseudoscalar mass $~m_{\pi}~$ and the 
condensate $~~\langle \psb \psi \rangle_{subt}~$. 
This work is in progress.

\pagebreak


\begin{thebibliography}{99}

\newcommand{\prd}[1]{Phys.~Rev.~{\bf D#1}\ }
\newcommand{\plb}[1]{Phys.~Lett.~{\bf #1B}\ }
\newcommand{\npb}[1]{Nucl.~Phys.~{\bf B#1}\ }
\newcommand{\prl}[1]{Phys.~Rev.~Lett.~{\bf #1}\ }
\newcommand{\pr}[1]{Phys.~Rep.~{\bf #1}\ }
\newcommand{\ap}[1]{Ann.~Phys.~{\bf #1}\ }
\newcommand{\cmp}[1]{Commun.~Math.~Phys.~{\bf #1}}
\newcommand{\rmp}[1]{Rev.~Mod.~Phys.~{\bf #1}}
\newcommand{\ptp}[1]{Prog.~Theor.~Phys.~{\bf #1}}
%
\bibitem{wil}   K. Wilson, \prd{10} (1974) 2445;
                in New phenomena in subnuclear physics, ed. A. Zichichi
                (Plenum, New York, 1977).
\bibitem{aoki}  S. Aoki, \prd{30} (1984) 2653; ~\prl{57} (1986) 3136;
                ~Phys. Lett. {\bf B190} (1987) 140.
\bibitem{cg1}   M. R. Hestenes and E. Stiefel, Journal of Research of the NBS
                {\bf 49} (1952) 409.
\bibitem{cg2}   M.~Engeli et al., {\it  Mitteilungen aus dem Institut f\"{u}r
                angewandte Mathematik} (Birkh\"{a}user Verlag,
                Berlin, 1959), Vol. 8, p. 24.
\bibitem{eig1}  Ph.~De Forcrand, A.~K\"onig, K.-H.~M\"utter,
                K.~Schilling and R.~Sommer,
                in: ~~Proc. Intern. Symp. on Lattice gauge theory
                (Brookhaven, 1986), (Plenum, New York, 1987).
\bibitem{eig2}   Ph.~De Forcrand, R.~Gupta, S.~G\"usken, K.-H.~M\"utter,
                A.~Patel, K.~Schilling and R.~Sommer,
                \plb{200} (1988) 143.
\bibitem{bkr}   K.~Bitar, A.~D.~Kennedy and P.~Rossi,
                Phys. Lett. {\bf B234} (1990) 333.
\bibitem{blls}  I. Barbour, E. Laermann, Th. Lippert and
                K. Schilling, Phys. Rev. {\bf D46} (1992) 3618.
\bibitem{qed1} A.~Hoferichter, V.K.~Mitrjushkin, M.~M\"uller\--Preus\-sker and
               Th.~Neuhaus,
               Nucl.~Phys.B (Proc.Suppl.)~{\bf 34} (1994) 537.
\bibitem{qed2} A.~Hoferichter, V.K.~Mitrjushkin, M.~M\"uller\--Preussker,
               Th.~Neuhaus and H.~St\"uben, 
               Nucl.Phys. {\bf B434} (1995) 358.
\bibitem{kawa}  N. Kawamoto, Nucl. Phys. {\bf B190} (1981) 617.
\bibitem{kasm}  N. Kawamoto and J. Smit, Nucl. Phys. {\bf B192} (1981) 100.
\bibitem{iiy}   S. Itoh, Y. Iwasaki and T. Yoshi\'{e},
                Phys. Rev. {\bf D36} (1987) 527.
\bibitem{uka}   A. Ukawa, CERN-TH-5245/88 (1988).
\bibitem{lues}  M. L\"uscher, Commun. Math. Phys. {\bf 54} (1977) 283.
\bibitem{nasi}  A.~Nakamura and R.~Sinclair, \plb{243} (1990) 396.
\bibitem{vdv}   B.L. Van Der Vaerden, {\it Mathematische
                Statistik}, Springer Verlag (1957).
\bibitem{deg1}  T.A. DeGrand, Comp.Phys.Comm. {\bf 52} (1988) 161.
\bibitem{deg2}  T.A. DeGrand and P. Rossi, Comp.Phys.Comm. {\bf 60} (1990) 211.
\bibitem{maian} M.~Bochicchio, L.~Maiani, G.~Martinelli, G.~Rossi and
                M.~Testa, \\
		    \npb{262} (1985) 331.
\bibitem{log1}  C. Bernard and M. Golterman, \prd{46} (1992) 853.
\bibitem{log2}  S. Sharpe, \prd{46} (1992) 3146.
\bibitem{log3}  R. Gupta, Nucl. Phys. {\bf B} (Proc. Suppl.)
                {\bf 42} (1994) 85.
\bibitem{lue1}  B. Bunk, K. Jansen, B. Jegerlehner, M. L\"uscher,
                H. Simma and  R. Sommer, Nucl. Phys. {\bf B} (Proc. Suppl.)
                {\bf 42} (1994) 49.
\end{thebibliography}
\end{document}